\documentclass[11pt,a4paper]{article}

\usepackage{jcappub,amsfonts,amsmath,amssymb,bm,graphicx}

\usepackage[active]{srcltx}

\title{Leptogenesis via hypermagnetic fields and baryon asymmetry}

\author[a,b]{Maxim Dvornikov}
\author[a]{Victor B. Semikoz}
\affiliation[a]{Pushkov Institute of the Terrestrial Magnetism,
the Ionosphere and Radiowave Propagation of the Russian Academy of Sciences, Troitsk, Moscow
Region, 142190, Russia}
\affiliation[b]{Institute of Physics, University of S\~{a}o Paulo, CP 66318, CEP 05315-970 S\~{a}o Paulo, SP, Brazil}
\emailAdd{maxim.dvornikov@usp.br}
\emailAdd{semikoz@yandex.ru}

\abstract{
We study baryon asymmetry generation originated from the
leptogenesis in the presence of hypermagnetic fields in the early
Universe plasma before the electroweak phase transition (EWPT). For the simplest
Chern-Simons (CS) wave configuration of hypermagnetic field we find the baryon asymmetry
growth when the hypermagnetic field value changes due to
$\alpha^2$-dynamo and the lepton asymmetry rises due to the
Abelian anomaly. We solve the corresponding integro-differential equations
for the lepton asymmetries describing such self-consistent dynamics for lepto- and
baryogenesis in the two scenarios : (i) when a primordial lepton asymmetry sits in
right electrons $e_R$; and (ii) when, in addition to $e_R$, a left lepton asymmetry for
$e_L$ and $\nu_{eL}$ arises due to chirality flip reactions provided by
inverse Higgs decays at the temperatures, $T<T_\mathrm{RL}\sim 10\thinspace\text{TeV}$.
We find that the baryon asymmetry of the Universe (BAU) rises very fast through such
leptogenesis, especially, in strong hypermagnetic fields. Varying (decreasing) the
CS wave number parameter $k_0<10^{-7}T_\mathrm{EW}$ one can recover the observable value of BAU,
$\eta_\mathrm{B}\sim 10^{-9}$, where $k_0=10^{-7}T_\mathrm{EW}$ corresponds to the maximum value for CS
wave number surviving ohmic dissipation of hypermagnetic field. In the scenario (ii)
one predicts the essential difference of the lepton numbers of right - and left
electrons at EWPT time,
$L_{eR} - L_{eL}\sim (\mu_{eR} - \mu_{eL})/T_\mathrm{EW}=\Delta \mu/T_\mathrm{EW}\simeq 10^{-5}$
that can be used as an initial condition for chiral asymmetry after EWPT.
}


\begin{document}

\maketitle

\section{Introduction}

It is well known that, at the high-temperature symmetric phase of the Standard Model (SM) all gauge bosons acquire a ``magnetic" mass gap $\sim g^2T$, except for the Abelian gauge field associated to the weak hypercharge. Such massless hypercharge field $Y_{\mu}$ in the hot universe plasma appears to be a progenitor of the Maxwellian field which evolves after ElectroWeak Phase Transition (EWPT). The hypercharge field, in its turn, may arise from phase transitions in the very early Universe, before EWPT, such as during the inflationary epoch \cite{Grasso:2000wj}.

In the absence of hypermagnetic fields
the baryon asymmetry of the Universe (BAU) can be produced
through a leptogenesis. In particular, such a case was considered in Ref.~\cite{Campbell:1992jd} assuming that an initial BAU,
having a negligible value in our situation, is preserved since it is stored in right electrons $e_R$. Then violation of the lepton number due to Abelian anomaly in a strong external hypercharge field provides the growth of lepton number from that value. Meanwhile lepton- and baryon number evolution proceeds preserving $B-L=\text{const}$. Thereby accounting for hypermagnetic fields one can enhance BAU as well. This allows us to assert (see also in Ref.~\cite{Giovannini:1997eg}) that fermion number ``sits" in hypermagnetic field.

One can easily understand why the authors of Ref.~\cite{Campbell:1992jd} considered the scenario of
BAU generation with one lepton generation chosen as $e_R$.
They found that the tiny Yukawa coupling to the
Higgs for right electrons, $h_e=\sqrt{2}m_e/v=2.94\times 10^{-6}$,
provides for such leptons the latest entering the equilibrium with
left particles through Higgs decays and inverse decays. This means
that any primordially-generated lepton number that occurred as
$e_R$ may not  be transformed into $e_L$ soon enough switching on
sphaleron interactions which wipe out the remaining BAU.

While other right-handed lepton (and all quark) species
enter the equilibrium with corresponding left particles even earlier,
e.g., because of $h_{\mu,\tau}\gg h_e$ for leptons, they are beyond
game for BAU generation if their partial asymmetries are zero from
the beginning.

On the other hand, in the presence of a nonzero right electron
chemical potential, $\mu_{eR}\neq 0$,
the Chern-Simons (CS) term $\sim\mu_{eR}\mathbf{B}_\mathrm{Y}\mathbf{Y}$ arises
in the effective Lagrangian for the hypercharge field $Y_{\mu}$
\cite{Giovannini:1997eg,Redlich:1984md,Semikoz:2011tm}. It modifies the
Maxwell equation in SM with parity violation producing additional pseudovector current
${\bf J}_5\sim \mu_{eR}{\bf B}_\mathrm{Y}$ in a plasma. It is this current which leads to the
important $\alpha_\mathrm{Y}$--helicity parameter in a modified Faraday equation. This parameter appears to be scalar.
We remind that the standard magnetohydrodynamic (MHD) parameter, which is generated by vortices in plasma, $\alpha_\mathrm{MHD} \sim - \langle {\bf v}\cdot (\nabla\times {\bf v}) \rangle /3$, is pseudoscalar. Obviously in
the isotropic early Universe such vortices are absent, at least,
at large scales we consider here.

The $\alpha^2_\mathrm{Y}$-dynamo amplification of a large-scale
hypermagnetic field~\cite{Semikoz:2009ye} for changing chemical
potential $\mu_{eR}(t)$ as well as its growth
$\partial_t\mu_{eR}>0$ due to the Abelian anomaly in the
self-consistent hypermagnetic field ${\bf B}_\mathrm{Y}$ were never
explored before in literature because of the difficulties to solve the
corresponding integro-differential equations. In the present work
we solve that problem in two scenarios.

First, in the case described in Sec.~\ref{SRB} as
saturation regime, we can avoid the adiabatic
approach $B_\mathrm{Y}(t)\approx \text{const}$ and $\mu_{eR}(t)\approx
\text{const}$~Ref. \cite{Giovannini:1997eg}, using exact solutions of the
integro-differential equations for $\partial_t\mu_{eR}\neq 0$ (Sec.~\ref{DEBAUHFBSR}).

At temperatures $T<T_\mathrm{RL}\simeq
10~\mathrm{TeV}$ chirality flip reactions enter the equilibrium since the rate of chirality flip processes, $\Gamma_\mathrm{RL}\sim T$,
becomes faster than the Hubble expansion $H\sim T^2$, $\Gamma_\mathrm{RL}> H$.
This motivates us to consider in the second scenario
the extended equilibrium state at $T<T_\mathrm{RL}$ when left leptons enter the
equilibrium with $e_R$ through inverse Higgs decays and acquire non-zero
asymmetries  $\sim \mu_{eL}(T)\equiv \mu_{\nu_{eL}}(T)\neq 0$
(Sec.~\ref{ATALELN}).

The left lepton asymmetries can grow from a negligible
(even zero) value at $T< T_\mathrm{RL}$
due to the corresponding Abelian anomaly
which has the opposite sign relatively to the sign of the
anomaly for $e_R$ and because left leptons have
different coupling constant $g' Y_\mathrm{L}/2$
with hypercharge field. Such a difference guarantees the presence of leptogenesis
in hypermagnetic fields even below $T_\mathrm{RL}$ all the way
down to $T_\mathrm{EW}$. Hence it supports generation of the BAU.

Note that for $T>T_\mathrm{RL}$, before left leptons enter the equilibrium with
right electrons, the anomaly for them was not efficient since the left
electron (neutrino) asymmetry was zero, $\mu_{eL}=\mu_{\nu_{eL}}=0$,
while a non-zero primordial right electron asymmetry, $\mu_{eR}\neq
0$, kept the baryon asymmetry at the necessary level.
In other words, for $T>T_\mathrm{RL}$ the Abelian anomaly for left particles
was present at the
stochastic level, with $\langle \delta j_L^{\mu} \rangle = 0 = \langle {\bf E}_\mathrm{Y}{\bf B}_\mathrm{Y} \rangle$
valid only on large scales.

Since $\mu_{eL}\equiv
\mu_{\nu_{eL}}\neq 0$ at temperatures $T<T_\mathrm{RL}$ there appear additional
macroscopic pseudovector currents in plasma ${\bf J}_5\sim \mu_{eL}{\bf B}_\mathrm{Y}$ which
modify $\alpha_\mathrm{Y}$--helicity parameter governing Maxwell equations for
hypermagnetic (hyperelectric) fields. This occurs due to the same polarization
effect described for $e_R$ in Ref.~\cite{Semikoz:2011tm} which led to the appearance of
CS term in the effective SM Lagrangian.

The kinetics of leptogenesis at temperatures $T_\mathrm{RL}>T>T_\mathrm{EW}$ is considered in
Sec.~\ref{KL}, where we give some details how to
solve the corresponding integro-differential equations.

In Sec.~\ref{DISC} we discuss our results illustrated in Figs.~\ref{yBeta1} and~\ref{yBeta2}
for the particular case of the CS wave configuration of hypermagnetic field.

\section{Saturated regime of baryogenesis}\label{SRB}

Let us follow the scenario in Ref.~\cite{Giovannini:1997eg} to discern
approximations used there and in our approach. There are five
chemical potentials in symmetric phase corresponding to
the five conserved numbers.
For the three global charges
$n_i=B/3 - L_i$ conserved in SM there are three chemical potentials $\mu_i$, $i=1,2,3$,
and $\mu_\mathrm{Y}$ is the fourth chemical potential which
corresponds to the conserved hypercharge commuting with the
SM Hamiltonian  (global $\langle Y \rangle = 0$). The fifth
chemical potential $\mu_{eR}\neq 0$ is the single {\it partial}
chemical potential for fermions corresponding to the right electron number
perturbatively conserved unless Abelian anomaly is taken into account.

Note that formally other {\it partial} fermion
chemical potentials are zero since there are no asymmetries for left
leptons including neutrinos, no partial quark asymmetries, etc.
There are relations between all five chemical potentials (see
Eq.~(2.16) in Ref.~\cite{Giovannini:1997eg}) given by the
conserved global charges including electroneutrality and
hypercharge neutrality of plasma $\langle Q \rangle = \langle Y
\rangle =0$.

Under such equilibrium conditions the last CS
anomaly term in the effective Lagrangian for the hypercharge field
$Y_{\mu}$ \cite{Giovannini:1997eg},
\begin{equation}\label{Lagrangian2}
  L = -\frac{1}{4}Y_{\mu\nu}Y^{\mu\nu}  - J_{\mu}Y^{\mu} +
  \frac{g'^2 \mu_{eR}}{4\pi^2}
  (\mathbf{B}_\mathrm{Y} \cdot \mathbf{Y}),
\end{equation}
originated by the one-loop diagram in FTFT~\cite{Redlich:1984md}
at the finite fermion density ($\mu_{eR}\neq 0$) in hot plasma
bath of the early Universe ($T>T_\mathrm{EW}$) \footnote{\label{footnt1}We changed sign
ahead the anomaly term in the SM Lagrangian comparing with what
authors had in Refs.~\cite{Semikoz:2009ye,Semikoz:2011tm}. This is because of the definition of
$\gamma_5$ matrix for right fermions in Ref.~\cite{Giovannini:1997eg} which the
authors~\cite{Semikoz:2009ye,Semikoz:2011tm} relied on. Such choice of signs in Refs.~\cite{Giovannini:1997eg,Semikoz:2009ye,Semikoz:2011tm}
corresponds to the
definition $\psi_R=(1-\gamma_5)\psi/2$. Note that the change
$\gamma_5\to - \gamma_5$ concerns also sign in the Abelian anomaly
in Eq.~\eqref{anomaly}. After such changes the latter
(Abelian anomaly) becomes exactly like in the book~\cite{Zee} (pg.~249)
where $\psi_R=(1+\gamma_5)\psi/2$. } leads to the appearance of
parity violation terms in the Lagrange equations for
hypermagnetic and hyperelectric fields. In particular, the
modified Maxwell equation takes the form,
\begin{equation}\label{Maxwell}
  -\frac{\partial \mathbf{E}_\mathrm{Y}}{\partial t} + \nabla\times \mathbf{B}_\mathrm{Y} =
  \left(
    \mathbf{J} + \frac{g'^2 \mu_{eR}}{ 4\pi^2}\mathbf{B}_\mathrm{Y}
  \right) =
  \sigma_\mathrm{cond}
  \left[
    \mathbf{E}_\mathrm{Y} + \mathbf{V}\times \mathbf{B}_\mathrm{Y} + \alpha_\mathrm{Y} \mathbf{B}_\mathrm{Y}
  \right],
\end{equation}
where the last pseudovector term originated by the CS term in
Eq.~(\ref{Lagrangian2}) is given by the hypermagnetic helicity
coefficent
\begin{equation}\label{alpha}
  \alpha_\mathrm{Y}(T)= \frac{g'^2 \mu_{eR}(T)}{4\pi^2\sigma_\mathrm{cond}(T)},
\end{equation}
that is scalar. In the Lagrangian~\eqref{Lagrangian2} and in
Maxwell equation~\eqref{Maxwell} the coefficient $g'=e/\cos \theta_W$ is the SM coupling, $J_{\mu}=(J_0, \mathbf{J})$ is
the vector (ohmic) current with zero time component due to the
electro-neutrality of the plasma as a whole, $J_0= \langle Q
\rangle = 0$. In Eq.~\eqref{Maxwell} we also used the Ohm law
$\mathbf{J}/\sigma_{\rm cond}=\mathbf{E}_\mathrm{Y} +
\mathbf{V}\times\mathbf{B}_\mathrm{Y}$ and introduced conductivity
$\sigma_\mathrm{cond}\sim 100~ T$ for the hot universe plasma.

Note that a more transparent way suggested in Ref.~\cite{Semikoz:2011tm} to get the CS term in Eq.~(\ref{Lagrangian2})
 allows to clarify its physical sense as the polarization effect
in an external hypermagnetic field.

Using the standard MHD approach one can neglect the displacement current $\partial {\bf
E}_\mathrm{Y}/\partial t$. Then we obtain from the
Maxwell equation (\ref{Maxwell}) the hyperelectric field $\mathbf{E}_\mathrm{Y}$,
\begin{equation}\label{electric}
  \mathbf{E}_\mathrm{Y}=-\mathbf{V}\times \mathbf{B}_\mathrm{Y} +
  \frac{\nabla\times \mathbf{B}_\mathrm{Y}}{\sigma_\mathrm{cond}} - \alpha_\mathrm{Y}\mathbf{B}_\mathrm{Y}.
\end{equation}
Now one can easily find the change of the CS number density
$n_\mathrm{CS}=g'^2(\mathbf{B}_\mathrm{Y} \cdot \mathbf{Y})/8\pi^2$,
\begin{equation}\label{CSdensity}
  \Delta n_\mathrm{CS} = \int_{t_0}^{t_\mathrm{EW}}\mathrm{d}t\frac{\mathrm{d}n_\mathrm{CS}}{\mathrm{d}t}=
  -\frac{g'^2}{4\pi^2}\int_{t_0}^{t_\mathrm{EW}}(\mathbf{E}_\mathrm{Y} \cdot \mathbf{B}_\mathrm{Y})\mathrm{d}t,
\end{equation}
that, in its turn, allows us to reveal the generation of BAU, $\eta_\mathrm{B}=(n_\mathrm{B} - n_{\bar{\mathrm{B}}})/s$, via
hypermagnetic fields,
\begin{equation}\label{baryon25}
  \eta_\mathrm{B}(t_\mathrm{EW})= - \left(\frac{3}{s}\right)\Delta n_\mathrm{CS}=
  \frac{3g'^2}{4\pi^2s}\int_{t_0}^{t_\mathrm{EW}}\left(\frac{(\nabla\times
  \mathbf{B}_\mathrm{Y})\cdot \mathbf{B}_\mathrm{Y}}{\sigma_\mathrm{cond}} -
  \alpha_\mathrm{Y}B_\mathrm{Y}^2\right)\mathrm{d}t.
\end{equation}
Here
$s=2\pi^2T^3g^*/45$ is the entropy density; $g^*=106.75$ is the number of relativistic degrees of freedom and we substituted in Eq.~\eqref{CSdensity} the hyperelectric
field from Eq.~\eqref{electric}.

So far we neither used the dependence $\mu_{eR}(t)$ on time nor exploited leptogenesis through hypercharge fields. To realize this idea
authors of Ref.~\cite{Giovannini:1997eg} suggested to account for the
Abelian anomaly for right electrons \footnote{The opposite (here
positive) sign in the Abelian anomaly comparing with what was used
in Ref.~\cite{Giovannini:1997eg} and in Refs.~\cite{Semikoz:2009ye,Semikoz:2011tm} corresponds
to the standard definition of $\gamma_5$-matrix as in the case of
analogous Adler anomaly in QED calculated in the book~\cite{Zee}.
See our footnote~\ref{footnt1} above.}
\begin{equation}\label{anomaly}
  \frac{\partial j^{\mu}_{R}}{\partial x^{\mu}} =
  + \frac{g'^2Y_\mathrm{R}^2}{64\pi^2}Y_{\mu\nu}\tilde{Y}^{\mu\nu}=
  + \frac{g'^2}{4\pi^2} (\mathbf{E}_\mathrm{Y} \cdot \mathbf{B}_\mathrm{Y}),
  \quad
  Y_\mathrm{R}=-2,
\end{equation}
that violates lepton number. In the uniform universe accounting for
the right electron (positron) asymmetry $n_{eR}-
n_{\bar{e}R}=\mu_{eR}T^2/6$ and adding the chirality flip
processes (inverse decays $e_R\bar{e}_L\to \varphi^{(0)}$,
$e_R\bar{\nu}_{eL}\to \varphi^{(-)}$) with the equivalent rates,
$\Gamma_\mathrm{RL}$ and $\Gamma=2\Gamma_\mathrm{RL}$, one finds from
Eq.~\eqref{anomaly}:
\begin{equation}\label{right}
  \frac{\partial \mu_{eR}}{\partial t} =
  \frac{6g'^2(\nabla\times \mathbf{B}_\mathrm{Y})\cdot\mathbf{B}_\mathrm{Y}}{4\pi^2T^2\sigma_\mathrm{cond}} -
  (\Gamma_\mathrm{B} + \Gamma)\mu_{eR},
\end{equation}
where we substituted the hyperelectric field from Eq.~\eqref{electric} and
neglected left electron (neutrino) abundance rising through
inverse Higgs decays. Direct Higgs decays do not contribute since
one assumes for simplicity the zero Higgs boson asymmetry,
$n_{\varphi}=n_{\bar{\varphi}}$.

The rate of all inverse Higgs decay processes $\Gamma_\mathrm{RL}$~\cite{Campbell:1992jd},
\begin{equation}\label{Gamma}
  \Gamma_\mathrm{RL} = 5.3\times 10^{-3}h_e^2
  \left(
    \frac{m_0}{T}
  \right)^2
  T =
  \left(
    \frac{\Gamma_0}{2t_\mathrm{EW}}
  \right)
  \left(
    \frac{1 -x}{\sqrt{x}}
  \right),
\end{equation}
vanishes just at EWPT time, $x=1$. Here $h_e=2.94\times 10^{-6}$
is the Yukawa coupling for electrons, $\Gamma_0=121$,
variable $x=t/t_\mathrm{EW}=(T_\mathrm{EW}/T)^2$ is given by the Friedman law, $m_0^2(T)=2DT^2(1-T_\mathrm{EW}^2/T^2)$ is the
temperature dependent effective Higgs mass at zero momentum and
zero Higgs vacuum expectation value. The coefficient $2D\approx 0.377$ for $m_0^2(T)$ is given by
the known masses of gauge bosons $m_Z$, $m_W$, the top quark mass
$m_t$ and a still problematic zero-temperature Higgs mass
estimated as $m_H\sim 125\thinspace\text{GeV}$ (see Ref.~\cite{Higgsmass}).

Note that the helicity term in Eq.~(\ref{electric}) proportional
to $\alpha_\mathrm{Y}\sim \mu_{eR}$ entered Eq.~(\ref{right})
as $-\Gamma_\mathrm{B}\ \mu_{eR}$ with the rate

\begin{equation*}
  \Gamma_\mathrm{B}=\frac{6(g'^2/4\pi^2)^2B_\mathrm{Y}^2}{T^2\sigma_\mathrm{cond}}.
\end{equation*}

Then in the {\it adiabatic approximation} $\dot{s}=\dot{T}=0$ and
$\partial \mu_{eR}/\partial t= 0$, we reproduce from Eq.~\eqref{right}
the right electron chemical potential that is similar to
Eq.~(2.26) in Ref.~\cite{Giovannini:1997eg} (with the opposite sign!)
\begin{equation*}
  \mu_{eR} \approx
  \frac{6g'^2(\nabla\times \mathbf{B}_\mathrm{Y})\cdot\mathbf{B}_\mathrm{Y}}
  {4\pi^2T^2\sigma_\mathrm{cond}(\Gamma + \Gamma_\mathrm{B})}.
\end{equation*}
Substituting $\alpha_\mathrm{Y}$ with such a chemical potential
into the integrand for the baryon asymmetry in
Eq.~\eqref{baryon25} we reproduce the result, cf.
Eq.~(2.32) in Ref.~\cite{Giovannini:1997eg},
\begin{align}\label{integral2}
  \eta_\mathrm{B}(t_\mathrm{EW}) = &
  \frac{3g'^2}{4\pi^2s}\int_{t_0}^{t_\mathrm{EW}}
  \left(
    1 - \frac{\Gamma_\mathrm{B}}{\Gamma + \Gamma_\mathrm{B}}
  \right)
  \frac{(\nabla\times \mathbf{B}_\mathrm{Y})\cdot\mathbf{B}_\mathrm{Y}}{\sigma_\mathrm{cond}}\mathrm{d}t
  \notag
  \\
  & \simeq
  \frac{3g'^2\Gamma[(\nabla\times \mathbf{B}_\mathrm{Y})\cdot\mathbf{B}_\mathrm{Y}]}
  {4\pi^2s(\Gamma + \Gamma_\mathrm{B})\sigma_\mathrm{cond}}
  \left(
    \frac{M_0}{2T_\mathrm{EW}^2}
  \right),
\end{align}
independently of the sign for $\gamma_5$ we used here as it should be. In
Eq.~\eqref{integral2} we put $\eta_\mathrm{B}(t_0)=0$ and used a saturated
value $B_\mathrm{Y}$ independent of time. Then assuming approximately
constant integrand we substituted the expansion time
$t_\mathrm{EW}=M_0/2T_\mathrm{EW}^2$,
$M_0=M_\mathrm{Pl}/1.66\sqrt{g^{*{}}}$.

In the scenario of Ref.~\cite{Giovannini:1997eg}, for the simplest
CS wave configuration of the hypercharge field, $Y_0 = Y_z =0$,
$Y_x=Y(t)\sin k_0z$, $Y_y=Y(t)\cos k_0z$, for which $(\nabla\times
\mathbf{B}_\mathrm{Y})\cdot\mathbf{B}_\mathrm{Y}=k_0B_\mathrm{Y}^2(t)$,
where $B_\mathrm{Y}(t)=k_0Y(t)$, with the use of the rates
$\Gamma$ and $\Gamma_\mathrm{B}$ defined above the integral in
first line in Eq.~\eqref{integral2} takes the form,
\begin{align}\label{integral5}
  \eta_\mathrm{B}(x=1) = & 7.3\times 10^{-4}
  \left(
    \frac{k_0}{10^{-7}T_\mathrm{EW}}
  \right)
  \notag
  \\
  & \times
  \int_{x_0}^1
  \frac{(1-x') x'^2 \mathrm{d}x'}{[1-x' + 0.16 x'^2 (B_\mathrm{Y}(x')/10^{20}\thinspace\text{G})^2]}
  \left(
    \frac{B_\mathrm{Y}(x')}{10^{20}\thinspace\text{G}}
  \right)^2.
\end{align}
Here the parameter $k_0/10^{-7}T_\mathrm{EW}\leq 1$ is given by the
maximum wave number surviving ohmic dissipation of hypermagnetic
fields at the finite conductivity $\sigma_\mathrm{cond}\simeq
100~T$.

For completeness we present here the $\alpha^2$-dynamo formula
from Ref.~\cite{Semikoz:2011tm} for amplification of the
hypermagnetic field in the Fourier representation,
$B_\mathrm{Y}(k,t)=B_\mathrm{Y}^{(0)}\exp
[\int_{t_0}^t\Bigl(\alpha_\mathrm{Y}(t')k -
\eta_\mathrm{Y}(t')k^2\Bigr)]\mathrm{d}t'$, with an arbitrary
scale $\Lambda\equiv
k^{-1}=\kappa\eta_\mathrm{Y}/\alpha_\mathrm{Y}$ and $\kappa>1$,
\begin{equation}\label{Faraday}
  B_\mathrm{Y}(x) = B_\mathrm{Y}^{(0)}\exp
  \left[
    42
    \left(
      \frac{1}{\kappa} - \frac{1}{\kappa^2}
    \right)
    \int_{x_0}^x\frac{\mathrm{d}x'}{\sqrt{x'}}
    \left(
      \frac{\xi_{eR}(x')}{10^{-4}}
    \right)^2
  \right].
\end{equation}
Here $\xi_{eR}=\mu_{eR}/T$ is the dimensionless right electron
chemical potential, $\eta_\mathrm{Y}=(\sigma_\mathrm{cond})^{-1}$
is the magnetic diffusion coefficient, and $x=t/t_\mathrm{EW}\leq
1$ is the dimensionless time used above. One can see from
Eq.~\eqref{integral5} that varying the two parameters: either the
seed hypermagnetic field $B_\mathrm{Y}^{(0)}$ in dynamo
formula~\eqref{Faraday} or the CS wave number $k_0$ (or both) one
can obtain the known BAU $\eta_\mathrm{B}(t_\mathrm{EW})\simeq 6\times
10^{-10}$.

\section{Dynamical evolution of BAU in hypermagnetic fields
beyond saturated regime}\label{DEBAUHFBSR}

For the lepton (baryon) asymmetry densities $L_a=n_a -
n_{\bar{a}}$ ($B=n_\mathrm{B}-n_{\bar{\mathrm{B}}}$) using Hooft's conservation law
for the electron generation,
\begin{equation}\label{Hooft}
  2\frac{{\rm d}L_{eL}}{\rm dt} + \frac{{\rm d}L_{eR}}{\rm dt} =
  \frac{1}{3}\frac{{\rm d}(n_\mathrm{B}-n_{\bar{\mathrm{B}}})/s}{\rm dt} =
  \frac{1}{3}\frac{{\rm d}\eta_\mathrm{B}}{\rm dt},
\end{equation}
assuming, as in Ref.~\cite{Giovannini:1997eg}, the absence of left
lepton asymmetries, $L_{eL}=0$, and the direct consequence of the
Abelian anomaly for right electrons~\eqref{anomaly},
$\mathrm{d}L_{eR}/\mathrm{d}t=(g'^2/4\pi^2s)(\mathbf{E}_\mathrm{Y}
\cdot \mathbf{B}_\mathrm{Y})$, we get BAU at the EWPT time
coinciding with Eq.~\eqref{baryon25} in the adiabatic case
$\dot{s}=0$,
\begin{equation}\label{integral}
  \eta_\mathrm{B}(t_\mathrm{EW})=\frac{3g'^2}{4\pi^2}\int_{t_0}^{t_\mathrm{EW}}
  \left(
    \frac{(\nabla\times \mathbf{B}_\mathrm{Y}) \cdot \mathbf{B}_\mathrm{Y}}
    {\sigma_\mathrm{cond}} - \alpha_\mathrm{Y} B_\mathrm{Y}^2
  \right)
  \frac{\mathrm{d}t}{s}.
\end{equation}
Let us use the notation $x=t/t_\mathrm{EW}=(T_\mathrm{EW}/T)^2$ taken from Friedman
law and substitute $L_{eR}=\xi_{eR}T^3/6s=y_\mathrm{R}T^3/(6s \times
10^4)$.

Again using the CS wave configuration with $(\nabla\times
\mathbf{B}_\mathrm{Y})\cdot
\mathbf{B}_\mathrm{Y}=k_0B_\mathrm{Y}^2(x)$ and relying on the
total kinetic equation for the right electron asymmetry in
Eq.~\eqref{right} where {\it contrary to Eq.~(\ref{integral5})
obtained from Eq.~(\ref{right}) in the saturation regime
$\partial_t\mu_{eR}(t)= 0$}, we get BAU in the case
$\partial_t\mu_{eR}(t)\neq 0$,
\begin{equation}\label{baryon}
  \eta_\mathrm{B}(x) = 1.14\times 10^{-4}\int_{x_0}^x \mathrm{d}x'
  \left[
    \frac{\mathrm{d}y_\mathrm{R}(x')}{\mathrm{d}x'} +
    \Gamma_0\frac{(1-x')}{\sqrt{x'}}y_\mathrm{R}(x')
  \right].
\end{equation}
To obtain Eq.~(\ref{baryon}) we have solved the kinetic equation~(\ref{right}) rewritten in new dimensionless notations in the
form
\begin{equation}\label{right2}
  \frac{\mathrm{d}y_\mathrm{R}}{\mathrm{d}x} =
  \left(
    B_0x^{1/2} - A_0y_\mathrm{R}
  \right)
  \left(
    \frac{B_\mathrm{Y}^{(0)}}{10^{20}\thinspace\text{G}}
  \right)^2
  x^{3/2}e^{\varphi (x)} - \Gamma_0\frac{(1-x)}{\sqrt{x}}y_\mathrm{R},
\end{equation}
where
\begin{equation*}
  B_0=6.4
  \left(
    \frac{k_0}{10^{-7}T_\mathrm{EW}}
  \right),
  \quad
  A_0=19.4,
\end{equation*}
are constants chosen for hypermagnetic fields normalized
\footnote{In numerical estimates we substitute the parameter
$k_0/(10^{-7}\ T_\mathrm{EW}) = 1$ that is the upper limit for the CS wave
number, $k_0\leq 10^{-7}T_\mathrm{EW}$, to avoid ohmic dissipation of
hypermagnetic field.}, the exponent $e^{\varphi}$ is given by the
hypermagnetic field squared,
\begin{equation*}
  e^{\varphi(x)} =
  \left(
    \frac{B_\mathrm{Y}(x)}{B_\mathrm{Y}^{(0)}}
  \right)^2,
\end{equation*}
and we substituted the hypermagnetic field in Eq.~(\ref{Faraday})
with the fastest growth, $\kappa=2$,
\begin{equation}\label{value}
  B_\mathrm{Y}(x) = B_\mathrm{Y}^{(0)} \exp
  \left[
    10.5\int_{x_0}^x\frac{y_\mathrm{R}^2(x')\mathrm{d}x'}{\sqrt{x'}}
  \right],
\end{equation}
that corresponds to the hypermagnetic field scale
$\Lambda=2\eta_\mathrm{Y}/\alpha_\mathrm{Y}$.
\begin{figure}
  \centering
  \includegraphics[scale=1.3]{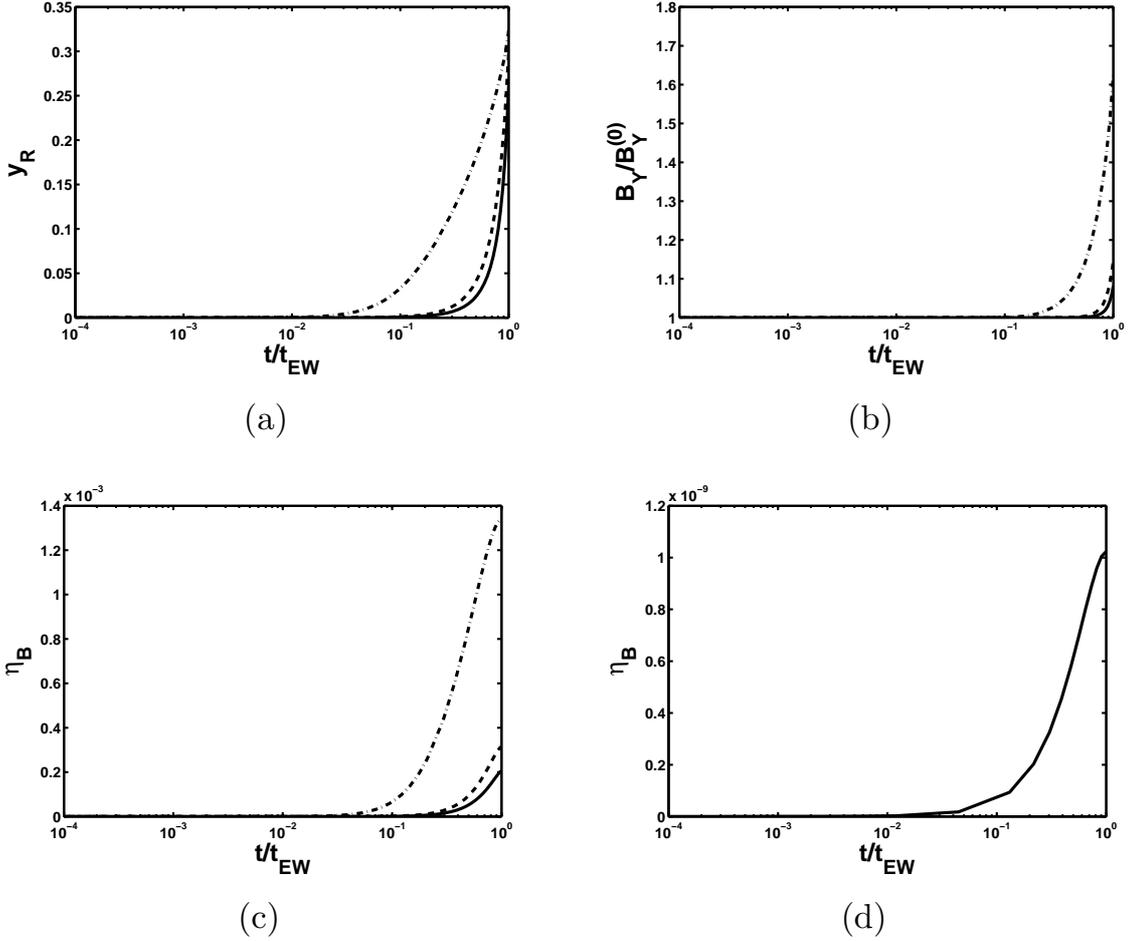}
  \caption{\label{yBeta1}
  Chern-Simons wave configuration for hypermagnetic field $B_\mathrm{Y}(t)$ with
  the maximum wave number $k_0/(10^{-7}\ T_\mathrm{EW}) = 1$
  surviving ohmic dissipation.
  (a) Normalized chemical potential $y_\mathrm{R} = 10^4 \times \xi_{eR}$ versus time.
  (b) Normalized hypermagnetic field versus time.
  (c) Baryon asymmetry versus time.
  The solid lines correspond to $B_\mathrm{Y}^{(0)} = 10^{20}\thinspace\text{G}$,
  the dashed lines to $B_\mathrm{Y}^{(0)} = 10^{21}\thinspace\text{G}$, and
  the dash-dotted lines to $B_\mathrm{Y}^{(0)} = 10^{22}\thinspace\text{G}$.
  (d) Baryon asymmetry versus time for the small wave number $k_0/(10^{-7}\ T_\mathrm{EW}) = 10^{-7}$ and $B_\mathrm{Y}^{(0)}=10^{20}\thinspace\text{G}$.}
\end{figure}

The most important step in the solution of the complicated
integro-differential  equation~\eqref{right2} is the following.
Accounting for the time derivative $\mathrm{d}e^{\varphi
(x)}/\mathrm{d}x=(21y_\mathrm{R}^2(x)/\sqrt{x})e^{\varphi (x)}$ with the exponent
$e^{\varphi (x)}$ given by Eq.~\eqref{right2} itself, or by the
first derivative $\mathrm{d}y_\mathrm{R}/\mathrm{d}x$,  let us differentiate
Eq.~\eqref{right2}. One obtains the non-linear differential
equation of the second order which we solve numerically,
\begin{align}\label{second}
  \frac{\mathrm{d}^2y_\mathrm{R}}{\mathrm{d}x^2} - & \frac{21y_\mathrm{R}^2}{\sqrt{x}}
  \left[
    \frac{\mathrm{d}y_\mathrm{R}}{\mathrm{d}x} + \Gamma_0\frac{(1-x)}{\sqrt{x}}y_\mathrm{R}
  \right]
  -\frac{(\mathrm{d}y_\mathrm{R}/\mathrm{d}x + \Gamma_0(1-x) y_\mathrm{R}/\sqrt{x})}{x^{3/2}(B_0x^{1/2} - A_0y_\mathrm{R})}
  \notag
  \\
  & \times
  \left(
    \frac{3}{2}B_0x -\frac{3}{2}A_0x^{1/2}y_\mathrm{R} -A_0x^{3/2}\frac{\mathrm{d}y_\mathrm{R}}{\mathrm{d}x}
  \right)
  \notag
  \\
  & +
  \Gamma_0\frac{(1-x)}{\sqrt{x}}\frac{\mathrm{d}y_\mathrm{R}}{\mathrm{d}x} -
  \Gamma_0\frac{(1 + x)}{2x^{3/2}}y_\mathrm{R}=0.
\end{align}
The initial conditions at $x_0=(T_\mathrm{EW}/T_\mathrm{RL})^2=10^{-4}$,
$T_\mathrm{EW}=100\thinspace\text{GeV}$, $T_\mathrm{RL}=10\thinspace\text{TeV}$,
are chosen as $y_\mathrm{R}(x_0)=10^{-6}$ (or $\xi_{eR}(x_0)=10^{-10}$) and
\begin{equation}
  \left.
    \frac{\mathrm{d}y_\mathrm{R}}{\mathrm{d}x}
  \right|_{x_0} =
  \left[
    B_0\sqrt{x_0}-A_0y_\mathrm{R}(x_0)
  \right]
  \left(
    \frac{B_\mathrm{Y}^{(0)}}{10^{20}\thinspace\text{G}}
  \right)^2x_0^{3/2} -
  \Gamma_0\frac{(1-x_0)}{\sqrt{x_0}}y_\mathrm{R}(x_0).
\end{equation}
The growth of BAU $\eta_\mathrm{B}(t)$ given by Eq.~\eqref{baryon} is
illustrated in Fig.~\ref{yBeta1}(c).

\section{Abelian (triangle) anomaly for left electrons and left neutrinos}\label{ATALELN}

In contrast to QED where the electron lepton number is conserved and
Abelian anomaly terms in an external electromagnetic field $F_{\mu\nu}$~\cite{Zee},
\begin{equation*}
  \frac{\partial j^{\mu }_L}{\partial x^{\mu}} =
  -\frac{e^2}{16\pi^2}F_{\mu\nu}\tilde{F}^{\mu\nu},
  \quad
  \frac{\partial j^{\mu }_R}{\partial x^{\mu}} =
  +\frac{e^2}{16\pi^2}F_{\mu\nu}\tilde{F}^{\mu\nu},
\end{equation*}
do not contradict to this law,
$\partial_{\mu}j^{\mu}=\partial_{\mu}(j_L^{\mu} + j_R^{\mu})=0$,
the leptogenesis in presence of an external hypercharge field
$Y_{\mu\nu}$ always exists, $\partial_{\mu}(j_L^{\mu} +
j_R^{\mu})\neq 0$,  since the coupling constants $e \to
g' Y_{R,L}/2$ are different for right singlet $e_R$ and left
doublet $L=(\nu_{eL}, e_L)^\mathrm{T}$.

In addition to Abelian anomaly for right electrons below we assume
the presence of Abelian anomalies for the left doublet at the
stage of chirality flip processes, $T<T_\mathrm{RL}$ \cite{Semikoz:2011tm},
\begin{equation}\label{left}
  \frac{\partial j^{\mu}_{L}}{\partial x^{\mu}} =
  - \frac{g'^2Y_\mathrm{L}^2}{64\pi^2}Y_{\mu\nu}\tilde{Y}^{\mu\nu} =
  - \frac{g'^2}{16\pi^2} (\mathbf{E}_\mathrm{Y} \cdot \mathbf{B}_\mathrm{Y}),
  \quad
  Y_\mathrm{L}=-1.
\end{equation}
These anomalies provide the appearance of the left electron
(neutrino) asymmetries. This means that the new CS terms in effective
Lagrangian with finite $\mu_{eL}\equiv\mu_{\nu_{eL}}\neq 0$ should
arise due to the polarization mechanism \cite{Semikoz:2011tm}.

Namely the 3-vector components $\mathbf{J}^Y_5$ for the
statistically averaged left electron and neutrino pseudocurrents,
\begin{equation*}
  J_{j5}^Y=\frac{f_L(g^{'})}{2}
  \left[
    \langle
      \bar{e}\gamma_j\gamma_5e
    \rangle +
    \langle
      \bar{\nu}_{eL}\gamma_j\gamma_5\nu_{eL}
    \rangle
  \right],
\end{equation*}
arise in plasma as additional macroscopic currents modifying
Maxwell equations,
\begin{equation}\label{gamma5}
  \mathbf{J}_5^Y =
  -\frac{g'^2(\mu_{eL} + \mu_{\nu_{eL}})}{16\pi^2}\mathbf{B}_\mathrm{Y}.
\end{equation}
Here the U(1) gauge coupling $f_L(g')=g' Y_\mathrm{L}/2$ plays the role of
an ``electric'' charge associated to $U_\mathrm{Y}(1)$, with
$Y_\mathrm{L}=-1$ being hypercharge of the left-handed electron (neutrino).

Obviously, $\mu_{eL}\neq \mu_{eR}$ in the scenarios
of Refs.~\cite{Campbell:1992jd,Giovannini:1997eg}, where $\mu_{eL}=0$ is kept
forever, while we assume that only as an initial condition,
$\mu_{eL}(t_0)=0$  starting at $T_0=T_\mathrm{RL}$.

If we accept our \emph{ansatz} with Abelian anomalies for left
particles~\eqref{left} we obtain from the same Hooft's
rule~\eqref{Hooft} the reduced coefficient in the expression
similar to Eq.~(\ref{integral}) $3g'^2/4\pi^2\to 3(g'^2/\pi^2)[1/4
-2\times (1/16)]=3g'^2/8\pi^2$ and from Eq.~(\ref{gamma5}) the
modified helicity coefficient $\alpha_\mathrm{Y}^\mathrm{mod}$ in
the hyperelectric field (\ref{electric}),
\begin{equation}\label{alphamod}
  \alpha_\mathrm{Y}^\mathrm{mod} = +\frac{g'^2(2\mu_{eR}+\mu_{eL})}{8\pi^2\sigma_\mathrm{cond}}.
\end{equation}
Finally the baryon asymmetry resulting from~\eqref{Hooft} takes the
form:
\begin{equation}\label{integral3}
  \eta_\mathrm{B}(t_\mathrm{EW}) = \frac{3g'^2}{8\pi^2}\int_{t_0}^{t_\mathrm{EW}}
  \left(
    \frac{(\nabla\times \mathbf{B}_\mathrm{Y})\cdot \mathbf{B}_\mathrm{Y}}{\sigma_\mathrm{cond}} -
    \alpha_\mathrm{Y}^\mathrm{mod}B_\mathrm{Y}^2
  \right)
  \frac{\mathrm{d}t}{s}.
\end{equation}
Note that due to Higgs boson inverse decays the chemical
potentials $\mu_{eR}(t)$ and $\mu_{eL}(t)$ entering to
$\alpha_\mathrm{Y}$ ($\alpha_\mathrm{Y}^\mathrm{mod}$) are changed
over time influencing also the amplitude of hypermagnetic field
$B_\mathrm{Y}(t)$ accordingly dynamo Eq.~\eqref{Faraday}.

\section{Kinetics of the leptogenesis}\label{KL}

Note that we consider the stage of chirality flip processes when
left electrons and left neutrinos supplied by the Higgs decays
(inversed decays) in the presence of right electrons enter the
equilibrium with the same densities at the temperatures below
$T<T_\mathrm{RL}$, $n_L=n_{eL}=n_{\nu_{eL}}$, having the same chemical
potentials $\mu_{eL}=\mu_{\nu_{eL}}$.

Accounting for the Abelian anomalies for $e_R$ and left doublet
$L = (\nu_{eL}, e_L)^\mathrm{T}$ we get the following system of kinetic
equations for the charged lepton asymmetry densities,
$n_R=n_{eR}-n_{\bar{e}R}$ and $n_L=n_{eL}-n_{\bar{e}L}$, or for the
corresponding lepton numbers, $L_{eL}=L_{\nu_{eL}}=n_L/s$ and
$L_{eR}=n_R/s$, \cite{Semikoz:2011tm}:
\begin{eqnarray}\label{system}
  &&\frac{{\rm d}L_{eR}}{\rm dt} =
  \frac{g'^2}{4\pi^2s}\mathbf{E}_\mathrm{Y} \cdot \mathbf{B}_\mathrm{Y} + 2\Gamma_\mathrm{RL}(L_{eL}-L_{eR}),
  \quad
  {\rm for}
  \quad
  e_R\bar{e}_L\to \varphi^{(0)}, e_R\bar{\nu}_{eL}\to \varphi^{(-)},\nonumber\\
  &&\frac{{\rm d}L_{eL}}{\rm dt} =
  -\frac{g'^2}{16\pi^2s}\mathbf{E}_\mathrm{Y} \cdot \mathbf{B}_\mathrm{Y} +\Gamma_\mathrm{RL}(L_{eR} - L_{eL}),
  \quad
  {\rm for}
  \quad
  \bar{e}_Re_L\to \bar{\varphi}^{(0)},\nonumber\\&&\frac{{\rm d}L_{\nu_{L}}}{\rm dt} =
  -\frac{g'^2}{16\pi^2s}\mathbf{E}_\mathrm{Y} \cdot \mathbf{B}_\mathrm{Y} +\Gamma_\mathrm{RL}(L_{eR} - L_{eL}),
  \quad
  {\rm  for}
  \quad
  \bar{e}_R\nu_{eL}\to \varphi^{(+)}.
\end{eqnarray}
Here the factor=2 in the
first line takes into account the equivalent reaction branches and
we neglected Higgs boson asymmetries,
$n_{\varphi}=n_{\bar{\varphi}}$, or the Higgs decays into leptons
do not contribute in kinetic equations~\eqref{system}.

One can easily see from the sum of equations~\eqref{system} that
in correspondence with BAU given by Eq.~(\ref{integral3}) the
inverse Higgs processes in the Hooft's rule
$\mathrm{d}\eta_\mathrm{B}/\mathrm{d}t = 3[\mathrm{d}L_{eR}/\mathrm{d}t + \mathrm{d}L_{eL}/\mathrm{d}t + \mathrm{d}L_{\nu_{eL}}/\mathrm{d}t]$
contribute via indirect way, through the hypermagnetic fields
only. Thus, baryon number sits in hypermagnetic field itself.

\begin{figure}
  \centering
  \includegraphics[scale=1.3]{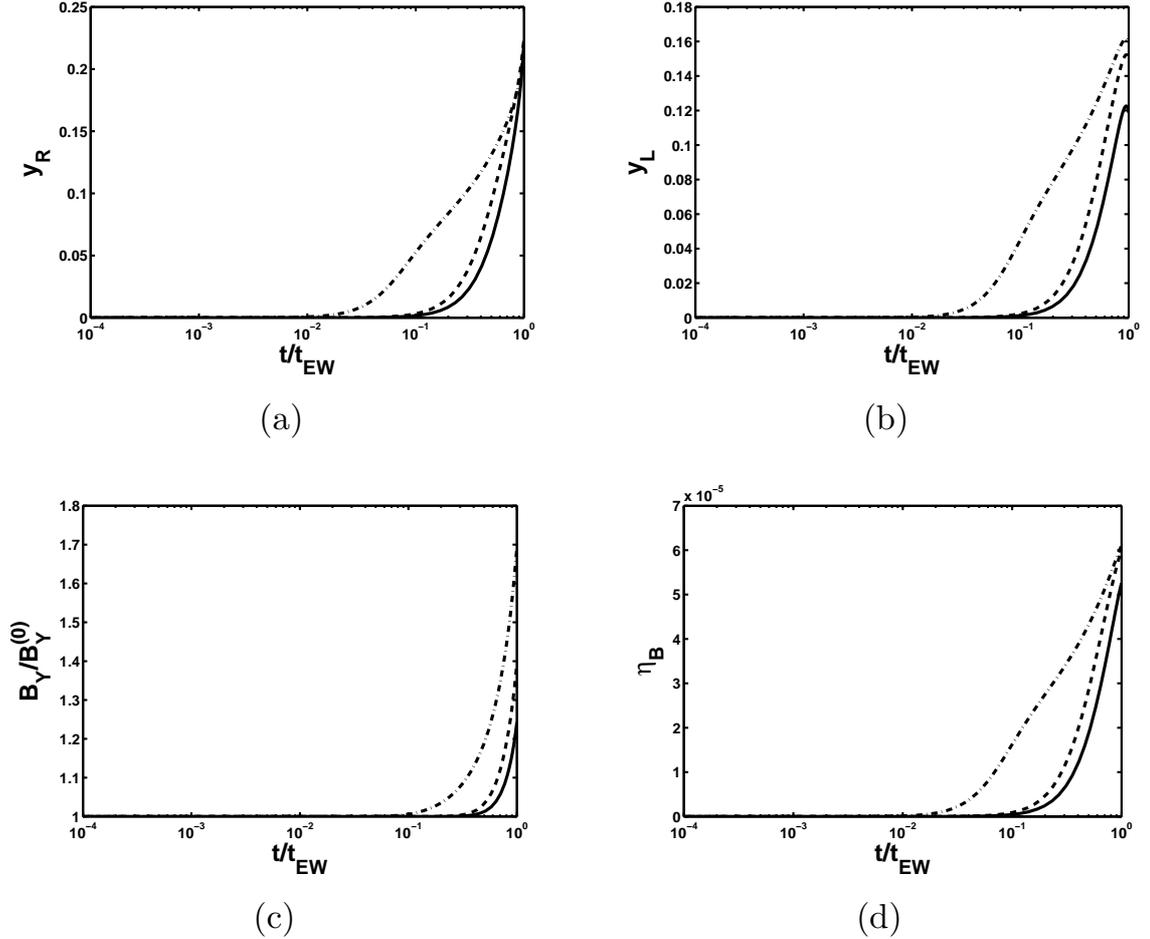}
  \caption{\label{yBeta2}
  Chern-Simons wave configuration for hypermagnetic field $B_\mathrm{Y}(t)$ with
  the maximum wave number $k_0/(10^{-7}\ T_\mathrm{EW})= 1$
  surviving ohmic dissipation.
  (a) Normalized chemical potential $y_\mathrm{R} = 10^4 \times \xi_{eR}$ versus time.
  (b) Normalized chemical potential $y_\mathrm{L} = 10^4 \times \xi_{eL}$ versus time.
  (c) Normalized hypermagnetic field versus time.
  (d) Baryon asymmetry versus time.
  The solid lines correspond to $B_\mathrm{Y}^{(0)} = 10^{20}\thinspace\text{G}$,
  the dashed lines to $B_\mathrm{Y}^{(0)} = 10^{21}\thinspace\text{G}$, and
  the dash-dotted lines to $B_\mathrm{Y}^{(0)} = 10^{22}\thinspace\text{G}$.}
\end{figure}

Below we write differential equations describing evolution of the right and
left asymmetries that follows from the kinetic equations~(\ref{system}).
We solved them numerically with the corresponding initial conditions and
the solutions are illustrated in Fig.~\ref{yBeta2}.

For right electrons $e_R$  we use the notation of the asymmetry
$y_\mathrm{R}(x)=10^4 \times \xi_{eR}(x)$ and get from the first equation in~(\ref{system})
after differentiation, cf. Eq.~(\ref{second}),
\begin{align}
  \frac{\mathrm{d}^2y_\mathrm{R}}{\mathrm{d}x^2} - & \frac{21(y_\mathrm{R}+y_\mathrm{L}/2)^2}{\sqrt{x}}
  \left[
    \frac{\mathrm{d}y_\mathrm{R}}{\mathrm{d}x} + \Gamma_0\frac{(1-x)}{\sqrt{x}}(y_\mathrm{R}-y_\mathrm{L})
  \right]
  \notag
  \\
  & -
  \frac{[\mathrm{d}y_\mathrm{R}/\mathrm{d}x +
  (\Gamma_0(1-x)/\sqrt{x})(y_\mathrm{R}-y_\mathrm{L})]}{x^{3/2}[B_0x^{1/2} - A_0(y_\mathrm{R}+y_\mathrm{L}/2)]}
  \notag
  \\
  & \times
  \left(
    \frac{3}{2}B_0x -\frac{3}{2}A_0x^{1/2}(y_\mathrm{R} + y_\mathrm{L}/2) -A_0x^{3/2}
    \left[
      \frac{\mathrm{d}y_\mathrm{R}}{\mathrm{d}x} + \frac{1}{2}\frac{\mathrm{d}y_\mathrm{L}}{\mathrm{d}x}
    \right]
  \right)
  \notag
  \\
  & +
  \Gamma_0\frac{(1-x)}{\sqrt{x}}
  \left(
    \frac{\mathrm{d}y_\mathrm{R}}{\mathrm{d}x}-\frac{\mathrm{d}y_\mathrm{L}}{\mathrm{d}x}
  \right) -
  \Gamma_0\frac{(1 + x)}{2x^{3/2}}(y_\mathrm{R} - y_\mathrm{L})=0.
\end{align}

For left leptons (both $e_L$ and $\nu_{eL}$) using the notation
$y_\mathrm{L}(x)=10^4 \times \xi_{eL}(x)$ one obtains from the second equation in (\ref{system}),

\begin{align}
  \frac{\mathrm{d}^2y_\mathrm{L}}{\mathrm{d}x^2} + & \frac{21(y_\mathrm{R}+y_\mathrm{L}/2)^2}{4\sqrt{x}}
  \left[
    \frac{\mathrm{d}y_\mathrm{R}}{\mathrm{d}x} + \Gamma_0\frac{(1-x)}{\sqrt{x}}(y_\mathrm{R}-y_\mathrm{L})
  \right]
  \notag
  \\
  & +
  \frac{1}{4}\frac{[\mathrm{d}y_\mathrm{R}/\mathrm{d}x +
  (\Gamma_0(1-x)/\sqrt{x})(y_\mathrm{R}-y_\mathrm{L})]}
  {x^{3/2}[B_0x^{1/2} - A_0(y_\mathrm{R}+y_\mathrm{L}/2)]}
  \notag
  \\
  & \times
  \left(
    \frac{3}{2}B_0x -\frac{3}{2}A_0x^{1/2}(y_\mathrm{R} + y_\mathrm{L}/2) -A_0x^{3/2}
    \left[
      \frac{\mathrm{d}y_\mathrm{R}}{\mathrm{d}x} +
      \frac{1}{2}\frac{\mathrm{d}y_\mathrm{L}}{\mathrm{d}x}
    \right]
  \right)
  \notag
  \\
  & +
  \Gamma_0\frac{(1-x)}{2\sqrt{x}}
  \left(
    \frac{\mathrm{d}y_\mathrm{L}}{\mathrm{d}x}-\frac{\mathrm{d}y_\mathrm{R}}{\mathrm{d}x}
  \right) -
  \Gamma_0\frac{(1 + x)}{4x^{3/2}}(y_\mathrm{L} - y_\mathrm{R})=0.
\end{align}

The initial conditions at $x_0=(T_\mathrm{EW}/T_\mathrm{RL})^2=10^{-4}$,
$T_\mathrm{EW}=100\thinspace\text{GeV}$, $T_\mathrm{RL}=10\thinspace\text{TeV}$,
are chosen as $y_\mathrm{R}(x_0)=10^{-6}$ (or $\xi_{eR}(x_0)=10^{-10}$) and
\begin{equation}
  \left.
    \frac{\mathrm{d}y_\mathrm{R}}{\mathrm{d}x}
  \right|_{x_0} =
  \left[
    B_0\sqrt{x_0}-A_0y_\mathrm{R}(x_0)
  \right]
  \left(
    \frac{B_\mathrm{Y}^{(0)}}{10^{20}\thinspace\text{G}}
  \right)^2x_0^{3/2} -
  \Gamma_0\frac{(1-x_0)}{\sqrt{x_0}}y_\mathrm{R}(x_0).
\end{equation}
For left particles we choose $y_\mathrm{L}(x_0)=\xi_{eL}(x_0)=0$ and
\begin{equation}\label{initial2}
  \left.
    \frac{\mathrm{d}y_\mathrm{L}}{\mathrm{d}x}
  \right|_{x_0} = -\frac{1}{4}
  \left[
    B_0\sqrt{x_0} -A_0y_\mathrm{R}(x_0)
  \right]
  \left(
    \frac{B_\mathrm{Y}^{(0)}}{10^{20}\thinspace\text{G}}
  \right)^2x_0^{3/2} +
  \frac{\Gamma_0}{2}\frac{(1-x_0)}{\sqrt{x_0}}y_\mathrm{R}(x_0).
\end{equation}

The BAU~(\ref{integral3}) for arbitrary $x$ takes the form,
\begin{equation}\label{integral4}
  \eta_\mathrm{B}(x)=0.57\times 10^{-4}
  \int_{x_0}^x \mathrm{d}x'
  \left\{
    \frac{\mathrm{d}y_\mathrm{R}(x')}{\mathrm{d}x'} + \Gamma_0\frac{(1-x')}{\sqrt{x'}}
    \left[
      y_\mathrm{R}(x') - y_\mathrm{L}(x')
    \right]
  \right\}.
\end{equation}

Accounting for the modified helicity parameter
$\alpha_\mathrm{Y}^\mathrm{mod}$ given by (\ref{alphamod}) the
$\alpha^2$-dynamo formula for the hypermagnetic field
$B_\mathrm{Y}(x)$ in the case of the fastest growth, see
Eq.~(\ref{value}), takes the form,
\begin{equation}\label{value2}
  B_\mathrm{Y}(x)=B_\mathrm{Y}^{(0)} \exp
  \left[
    10.5\int_{x_0}^x\frac{[y_\mathrm{R}(x') + y_\mathrm{L}(x')/2]^2\mathrm{d}x'}{\sqrt{x'}}
  \right].
\end{equation}

\section{Discussion}\label{DISC}

We study BAU evolution in the electroweak plasma of the early
Universe in the epoch before EWPT in the presence of a large-scale
hypermagnetic field $\mathbf{B}_\mathrm{Y}$ and assuming a primordial
right electron asymmetry $\mu_{eR}\neq 0$ that grows in a hypercharge field
due to the Abelian anomaly. We considered dynamical BAU evolution
in the two scenarios: (i) neglecting left lepton asymmetry,
$\mu_{eL}=y_\mathrm{L}=0$ (Sec.~3); and (ii) assuming its appearance,
$\mu_{eL}\neq 0$, at temperatures below the chirality flip
processes enter equilibrium, $T<T_\mathrm{RL}\sim
10\thinspace\text{TeV}$ (Secs.~\ref{ATALELN} and~\ref{KL}).

In the first scenario such a violation of the lepton number given
by Eq.~(\ref{anomaly}) leads to the growth of BAU shown in
Fig~\ref{yBeta1}(c) up to the big value $\eta_\mathrm{B}\sim
10^{-3}-10^{-4}$. We obtained that estimate for the largest CS
wave number $k_0=10^{-7}T_\mathrm{EW}$ surviving ohmic dissipation
in the case of simplest CS wave configuration for
$\mathbf{B}_\mathrm{Y}$. Note that decreasing $k_0$ down to
$k_0=(10^{-13}-10^{-14})T_\mathrm{EW}$, we can get the known BAU
value $\eta_\mathrm{B}\simeq 6\times10^{-10}$ for large scales of
hypermagnetic field, cf. Fig.~\ref{yBeta1}(d).

The maximum right electron asymmetry at the EWPT time,
$y_\mathrm{R}(t_\mathrm{EW})\simeq 0.3$,  is not sufficient to amplify the
seed hypermagnetic field $B_\mathrm{Y}^{(0)}$ too much. One can
easily estimate from the dynamo formula (\ref{value}) that even in
the case the fastest growth the hypermagnetic field $B
_\mathrm{Y}(x)$ rises a few times only, cf. Fig~\ref{yBeta1}(a).
This is a consequence of our choice of the simplest CS wave
configuration of $\mathbf{B}_\mathrm{Y}$. We think that for the
3D--configuration of hypermagnetic field the dynamo amplification
should be stronger $n^2$--times, where $n=\pm 1, \pm 2,\dots$ is the
number of knots of the tangled $\mathbf{B}_\mathrm{Y}$-field.
Moreover, for the chosen CS wave configuration a seed field
$B_\mathrm{Y}^{(0)}$ should be strong from the beginning to
influence the growth of the lepton asymmetry $y_\mathrm{R}$ and BAU,
$\eta_\mathrm{B}$. The stronger $B_\mathrm{Y}^{(0)}$  the more efficient
leptogenesis is (see in Fig.~\ref{yBeta1}).

We conclude that for non-helical CS-wave
one needs a preliminary amplification of $B_\mathrm{Y}^{(0)}$ in epoch
of inflation to provide both dynamo mechanism and
baryogenesis in our causal scenario in radiation era.

In the more realistic second scenario involving left electrons $e_L$ and neutrinos $\nu_{eL}$ we get the growth of their asymmetries $y_\mathrm{L}(x)$ due to the corresponding Abelian anomalies that become efficient at temperatures $T< T_\mathrm{RL}$ when inverse Higgs decays provide chirality flip $e_R\to L=(\nu_{eL}, e_L)^\mathrm{T}$ (Sec.~\ref{ATALELN}). We reveal the additional CS term in effective Lagrangian for the hypercharge field $Y_{\mu}$ arising due to polarization mechanism~\cite{Semikoz:2011tm} that modifies the hypermagnetic helicity coefficient in Eq.~(\ref{alphamod}) combined from the right and left electron asymmetries. As a result the growth of $y_\mathrm{R}(x)$ and $\eta_\mathrm{B}(x)$ reduces comparing with the first scenario for $\mu_{eR}\neq 0$ alone (see in Fig.~\ref{yBeta2}).

The important issue in such scenario is a non-zero difference of lepton numbers, $L_{eR} - L_{eL}\neq 0$ at the EWPT time, that can be used as a possible starting value for chiral anomaly which provides evolution of Maxwellian fields down to temperatures $T\sim 10\thinspace\text{MeV}$~\cite{Boyarsky:2011uy}. The corresponding difference of asymmetries seen in our Fig.~\ref{yBeta2}(a,b) is estimated at the level  $\xi_{eR}-\xi_{eL}=\mid \Delta \mu/T_\mathrm{EW}\mid \simeq 10^{-5}$ for $B_0^Y=10^{20}~G$ that is a bit below values $\Delta \mu/T_\mathrm{EW}$ seen in Fig.~1 in Ref.~\cite{Boyarsky:2011uy}.

We missed here the sphaleron wash-out of BAU due to the appearance of left lepton asymmetries believing as in Ref.~\cite{Campbell:1992jd} that time before EWPT during cooling $T_\mathrm{RL}>T>T_\mathrm{EW}$ is short
to evolve such process. We plan to study that effect in future as well as to generalize our approach for the case of 3D-configuration of hypermagnetic field.

\acknowledgments

We acknowledge Jose Valle and Dmitry Sokoloff for fruitful discussions. MD is thankful to FAPESP (Brazil) for a grant.

\end{document}


\maketitle


There are two kinds of mistakes in our previous work~\cite{Dvornikov:2011ey}. The values of hypermagnetic field in eqs.~(3.5) and~(5.7) coming from eq.~(2.12) were not appropriate for the case of Chern-Simons (CS) wave and the values of baryon asymmetry in eqs.~(3.3) and~(5.6) were overestimated due to the incorrect coefficient.

The $\alpha^2$-dynamo formula~(2.12) is applicable for an arbitrary 1D-configuration of hypermagnetic fields with the wave number $k\simeq \alpha_\mathrm{Y}/\eta_\mathrm{Y}$ that, in turn, changes over time through the lepton asymmetry,  $k(t)\sim \alpha_\mathrm{Y}(t)\sim \mu_{e_\mathrm{R}}(t)$. However, assuming such simplest hypercharge field configuration as the CS wave, $Y_0=Y_z=0$, $Y_x=Y(t)\sin k_0z$, $Y_y=Y(t)\cos k_0z$, {\it with the fixed wave number} $k_0=\mathrm{const}$, we should use the exact solution of Faraday equation
\begin{equation}
  B_\mathrm{Y}(t)=B_\mathrm{Y}^{(0)}
  \exp
  \left\{
    k_0\int_{t_0}^t[\alpha_\mathrm{Y}(t') - k_0\eta_\mathrm{Y}(t')]\mathrm{d}t'
  \right\},
  \tag{1}
\end{equation}
instead of eq.~(2.12). Therefore instead of eq.~(3.5) in ref.~\cite{Dvornikov:2011ey}, substituting $\alpha_\mathrm{Y}$ given by eq.~(2.3), one gets
\begin{equation}\label{ByR}
  B_\mathrm{Y}(t) =
  B_\mathrm{Y}^{(0)}
  \exp
  \left\{
    3.5
    \left(
      \frac{k_0}{10^{-7}T_\mathrm{EW}}
    \right)
    \int_{x_0}^x
    \left[
      \frac{y_\mathrm{R}(x')}{\pi} - 0.1
      \left(
        \frac{k_0}{10^{-7}T_\mathrm{EW}}
      \right)\sqrt{x'}
    \right]{\rm d}x'
  \right\},
  \tag{2}
\end{equation}
and instead of eq.~(5.7) substituting $\alpha^\mathrm{mod}_\mathrm{Y}$ given by eq.~(4.3) we obtain
\begin{align}\label{ByRyL}
  B_\mathrm{Y}(t) = & B_\mathrm{Y}^{(0)}\exp
  \bigg\{
    3.5
    \left(
      \frac{k_0}{10^{-7}T_\mathrm{EW}}
    \right)
    \notag
    \\
    & \times
    \int_{x_0}^x
    \left[
      \frac{(y_\mathrm{R}(x')+ y_\mathrm{L}(x')/2)}{\pi} - 0.1
      \left(
        \frac{k_0}{10^{-7}T_\mathrm{EW}}
      \right)
      \sqrt{x'}
    \right]
    {\rm d}x'
  \bigg\},
  \tag{3}
\end{align}
where $x=t/t_\mathrm{EW}\leq 1$ is the dimensionless time and $y_\mathrm{R,L}=10^4(\mu_{e_\mathrm{R,L}}/T)\equiv 10^4\xi_{e_\mathrm{R,L}}$ are the lepton asymmetries.
One can see from eqs.~(\ref{ByR}) and~(\ref{ByRyL}) that dynamo amplification occurs negligible  even for the maximum wave number $k_0=10^{-7}T_\mathrm{EW}$ for which hypermagnetic field survives against ohmic dissipation. Thus, figures~1(b) and~2(c) in ref.~\cite{Dvornikov:2011ey} for $B_\mathrm{Y}/B_\mathrm{Y}^{(0)}$, based on incorrect eqs.~(3.5) and~(5.7), were false and we do not replot them here. As a result, other plots are almost insensitive to the value $B_\mathrm{Y}^{(0)}$. Using eq.~(\ref{ByR}) and differentiating eq.~(3.4) we get, instead of eq.~(3.6), the following non-linear second order differential equation, which we solve numerically,
\begin{align}\label{second}
  \frac{\mathrm{d}^2y_\mathrm{R}}{\mathrm{d}x^2} - & B_0
  \Bigl(
    3.48 \times 10^{-1} y_\mathrm{R} - 1.7\times 10^{-2} \times B_0 \sqrt{x}
  \Bigr)
  \left[
    \frac{\mathrm{d}y_\mathrm{R}}{\mathrm{d}x} + \Gamma_0\frac{(1-x)}{\sqrt{x}}y_\mathrm{R}
  \right]
  \notag
  \\
  & - \frac{(\mathrm{d}y_\mathrm{R}/\mathrm{d}x + \Gamma_0(1-x) y_\mathrm{R}/\sqrt{x})}{x^{3/2}(B_0x^{1/2} - A_0y_\mathrm{R})}
  \left(
    2B_0x -\frac{3}{2}A_0x^{1/2}y_\mathrm{R} -A_0x^{3/2}\frac{\mathrm{d}y_\mathrm{R}}{\mathrm{d}x}
  \right)
  \notag
  \\
  & +
  \Gamma_0\frac{(1-x)}{\sqrt{x}}\frac{\mathrm{d}y_\mathrm{R}}{\mathrm{d}x} -
  \Gamma_0\frac{(1 + x)}{2x^{3/2}}y_\mathrm{R}=0.
  \tag{4}
\end{align}
The initial conditions for this equation are correct including eq.~(3.7).

Using eq.~(\ref{ByRyL}) we get, instead of eqs.~(5.2) and~(5.3),
\begin{align}\label{second2}
  \frac{\mathrm{d}^2y_\mathrm{R}}{\mathrm{d}x^2} - & B_0
  \Bigl(
    3.48 \times 10^{-1} \times (y_\mathrm{R} + y_\mathrm{L}/2) - 1.7\times 10^{-2} \times B_0 \sqrt{x}
  \Bigr)
  \notag
  \\
  & \times
  \left[
    \frac{\mathrm{d}y_\mathrm{R}}{\mathrm{d}x} + \Gamma_0\frac{(1-x)}{\sqrt{x}}(y_\mathrm{R}-y_\mathrm{L})
  \right]
  \notag
  \\
  & -
  \frac{[\mathrm{d}y_\mathrm{R}/\mathrm{d}x +
  (\Gamma_0(1-x)/\sqrt{x})(y_\mathrm{R}-y_\mathrm{L})]}{x^{3/2}[B_0x^{1/2} - A_0(y_\mathrm{R}+y_\mathrm{L}/2)]}
  \notag
  \\
  & \times
  \left(
    2B_0x -\frac{3}{2}A_0x^{1/2}(y_\mathrm{R} + y_\mathrm{L}/2) -A_0x^{3/2}
    \left[
      \frac{\mathrm{d}y_\mathrm{R}}{\mathrm{d}x} + \frac{1}{2}\frac{\mathrm{d}y_\mathrm{L}}{\mathrm{d}x}
    \right]
  \right)
  \notag
  \\
  & +
  \Gamma_0\frac{(1-x)}{\sqrt{x}}
  \left(
    \frac{\mathrm{d}y_\mathrm{R}}{\mathrm{d}x}-\frac{\mathrm{d}y_\mathrm{L}}{\mathrm{d}x}
  \right) -
  \Gamma_0\frac{(1 + x)}{2x^{3/2}}(y_\mathrm{R} - y_\mathrm{L})=0,
  \tag{5}
  \\
  \label{second2a}
  \frac{\mathrm{d}^2y_\mathrm{L}}{\mathrm{d}x^2} + & \frac{B_0}{4}
  \Bigl(
    3.48 \times 10^{-1} \times (y_\mathrm{R} + y_\mathrm{L}/2) - 1.7\times 10^{-2} \times B_0 \sqrt{x}
  \Bigr)
  \notag
  \\
  & \times
  \left[
    \frac{\mathrm{d}y_\mathrm{R}}{\mathrm{d}x} + \Gamma_0\frac{(1-x)}{\sqrt{x}}(y_\mathrm{R}-y_\mathrm{L})
  \right]
  \notag
  \\
  & +
  \frac{1}{4}\frac{[\mathrm{d}y_\mathrm{R}/\mathrm{d}x +
  (\Gamma_0(1-x)/\sqrt{x})(y_\mathrm{R}-y_\mathrm{L})]}
  {x^{3/2}[B_0x^{1/2} - A_0(y_\mathrm{R}+y_\mathrm{L}/2)]}
  \notag
  \\
  & \times
  \left(2B_0x -\frac{3}{2}A_0x^{1/2}(y_\mathrm{R} + y_\mathrm{L}/2) -A_0x^{3/2}
    \left[
      \frac{\mathrm{d}y_\mathrm{R}}{\mathrm{d}x} +
      \frac{1}{2}\frac{\mathrm{d}y_\mathrm{L}}{\mathrm{d}x}
    \right]
  \right)
  \notag
  \\
  & +
  \Gamma_0\frac{(1-x)}{2\sqrt{x}}
  \left(
    \frac{\mathrm{d}y_\mathrm{L}}{\mathrm{d}x}-\frac{\mathrm{d}y_\mathrm{R}}{\mathrm{d}x}
  \right) -
  \Gamma_0\frac{(1 + x)}{4x^{3/2}}(y_\mathrm{L} - y_\mathrm{R})=0,
  \tag{6}
\end{align}
obeying the same initial conditions~(5.4) and~(5.5). We show in figure~\ref{plotyRL}(a,b) the evolution of lepton asymmetries $y_\mathrm{R,L}(x)$ given by solutions of eqs.~(\ref{second2}) and~(\ref{second2a}) instead of figures~2(a,b) in ref.~\cite{Dvornikov:2011ey}.
\begin{figure}
  \centering
  \includegraphics[scale=1.3]{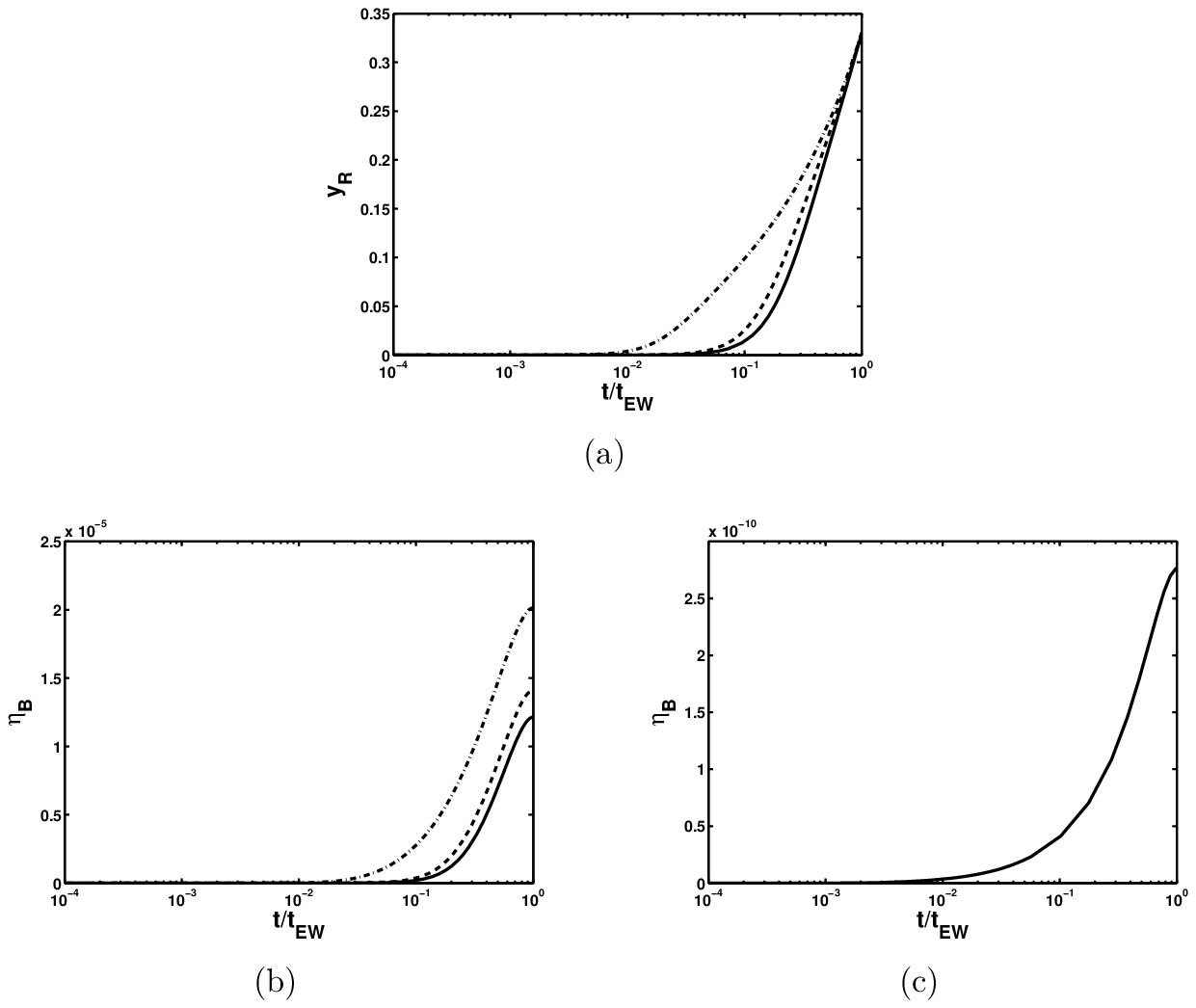}
  \caption{\label{plotyR}
  (a) Normalized chemical potential $y_\mathrm{R} = 10^4 \times \xi_{e_\mathrm{R}}$ versus time.
  (b) Baryon asymmetry versus time.
  Panels~(a) and~(b) are built for $k_0/(10^{-7}\ T_\mathrm{EW}) = 1$.
  The solid lines correspond to $B_\mathrm{Y}^{(0)} = 10^{20}\thinspace\text{G}$,
  the dashed lines to $B_\mathrm{Y}^{(0)} = 10^{21}\thinspace\text{G}$, and
  the dash-dotted lines to $B_\mathrm{Y}^{(0)} = 10^{22}\thinspace\text{G}$.
  (d) Baryon asymmetry versus time for the small wave number $k_0/(10^{-7}\ T_\mathrm{EW}) = 10^{-5}$ and $B_\mathrm{Y}^{(0)}=10^{20}\thinspace\text{G}$.}
\end{figure}
\begin{figure}
  \centering
  \includegraphics[scale=1.3]{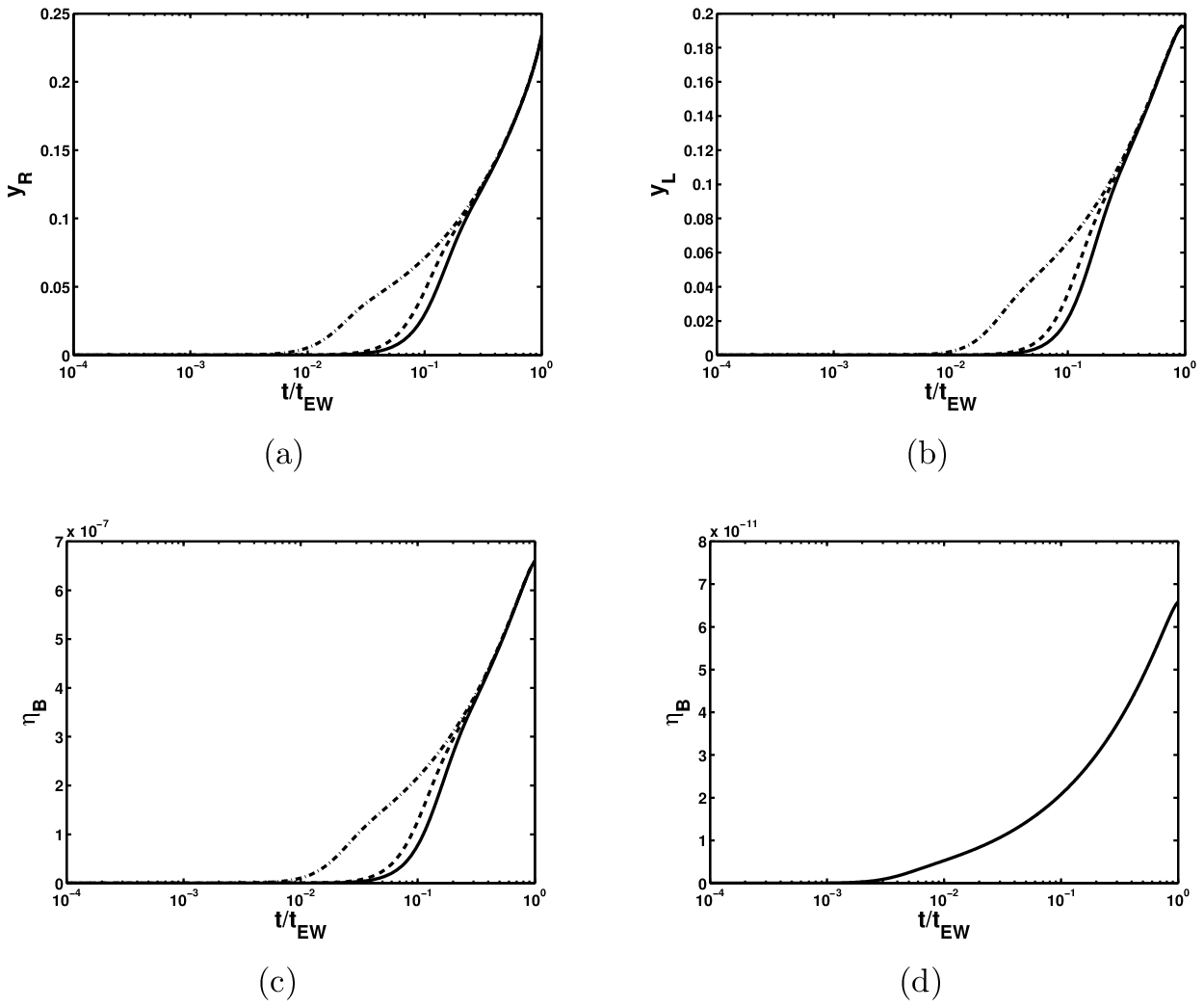}
  \caption{\label{plotyRL}
  (a) Normalized chemical potential $y_\mathrm{R} = 10^4 \times \xi_{e_\mathrm{R}}$ versus time.
  (b) Normalized chemical potential $y_\mathrm{L} = 10^4 \times \xi_{e_\mathrm{L}}$ versus time.
  (c) Baryon asymmetry versus time.
  Panels~(a)-(c) are built for $k_0/(10^{-7}\ T_\mathrm{EW}) = 1$.
  The solid lines correspond to $B_\mathrm{Y}^{(0)} = 10^{20}\thinspace\text{G}$,
  the dashed lines to $B_\mathrm{Y}^{(0)} = 10^{21}\thinspace\text{G}$, and
  the dash-dotted lines to $B_\mathrm{Y}^{(0)} = 10^{22}\thinspace\text{G}$.
  (d) Baryon asymmetry versus time for the small wave number $k_0/(10^{-7}\ T_\mathrm{EW}) = 10^{-4}$ and $B_\mathrm{Y}^{(0)}=10^{20}\thinspace\text{G}$.}
\end{figure}

We correct also wrong values of the baryon asymmetries (BAU) in eqs.~(3.3) and~(5.6) in ref.~\cite{Dvornikov:2011ey}, where the number of relativistic degrees of freedom, $g^*=106.75$, was not accounted for in the denominator.
It led to the incorrect plots in figure~1(c,d) and in figure~2(d) correspondingly. The correct expressions read
\begin{equation}\label{3.3}
  \eta_\mathrm{B}(x) = 1.068\times 10^{-6}
  \int_{x_0}^x{\rm d}x'
  \left\{
    \frac{{\rm d}y_\mathrm{R}(x')}{{\rm{d}}x'} + \Gamma_0\frac{(1 - x')}{\sqrt{x'}}y_\mathrm{R}(x')
  \right\},
  \tag{7}
\end{equation}
instead of eq. (3.3) in scenario with right electron asymmetry $y_\mathrm{R}=10^4 \times (\mu_{e_\mathrm{R}}/T)$ evolution and
\begin{equation}\label{5.6}
  \eta_\mathrm{B}(x)=5.34\times 10^{-7}
  \int_{x_0}^x{\rm d}x'
  \left\{
    \frac{{\rm d}y_\mathrm{R}(x')}{{\rm{d}}x'} + \Gamma_0\frac{(1 - x')}{\sqrt{x'}}
    \left[
      y_\mathrm{R}(x') - y_\mathrm{L}(x')
    \right]
  \right\},
  \tag{8}
\end{equation}
instead of eq.~(5.6) in scenario when both lepton asymmetries $y_\mathrm{R}$ and $y_\mathrm{L}=10^4 \times (\mu_{e_\mathrm{L}}/T)$ evolve.

Below using the numerical solution of eq.~(\ref{second}) for $y_\mathrm{R}(x)$ shown here in figure~\ref{plotyR}(a) we replot figure~1(c) in ref.~\cite{Dvornikov:2011ey} as figure~1(b) given by eq.~(\ref{3.3}) for the same maximum wave number of the Chern-Simons wave $k_0=10^{-7}T$. We replot also figure~1(d) in ref.~\cite{Dvornikov:2011ey} as new figure~\ref{plotyR}(c) given by eq.~(\ref{3.3}) changing $k_0=10^{-14}T$ used in ref.~\cite{Dvornikov:2011ey} to $k_0=10^{-12}T$. Such wave number provides observable BAU at $t=t_\mathrm{EW}$, $\eta_\mathrm{B}\sim 10^{-10}$.

We also present BAU evolution in figure~\ref{plotyRL}(c,d) instead of figure~2(d) in ref.~\cite{Dvornikov:2011ey} using here the correct eq.~(\ref{5.6}) based on the solution of eqs.~(\ref{second2}) and~(\ref{second2a}) for asymmetries $y_\mathrm{R,L}(x)$. The curves in figure~\ref{plotyRL}(c) are plotted for the same $k_0=10^{-7}T$ and  in figure~\ref{plotyRL}(d) for $k_0=10^{-11}T$.

\acknowledgments

One of the authors (MD) is thankful to FAPESP (Brazil) for a grant.